\documentclass[12pt]{article}

\usepackage{latexsym}
\usepackage{amsfonts}
\usepackage{amssymb}
\usepackage{graphics}
\usepackage{graphicx}
\usepackage{epsf}
\def\){\right)}
\def\({\left( }
\def\]{\right] }
\def\[{\left[ }
\def\half{\frac{1}{2}}
\def\d{\partial}

\def\be{\begin{equation}}
\def\ee{\end{equation}}
\def\ba{\begin{eqnarray}}
\def\ea{\end{eqnarray}}
\def\no{\nonumber \\}

\def\Z{{\mathbb{Z}}}
\def\C{{\mathbb{C}}}
\newcommand{\ket}{\rangle}
\newcommand{\bra}{\langle}
\def\ra{\rangle}
\def\la{\langle}

\begin{document}
\title{Closed string tachyon potential and $tt^*$ equation}

\author {Sunggeun Lee and Sang-Jin Sin \\
{\em Department of Physics, Hanyang University, 133-791, Seoul}
}

\maketitle


\begin{abstract}
Recently Dabholkar and Vafa proposed that  closed string
tachyon potential for non-supersymmetric orbifold $\C/\Z_3$ in terms of the solution of a $tt^*$ equation. We extend this result  to $\C^2/\Z_n$  for $n=3,4,5$. Interestingly, the tachyon potentials for $n=3$ and 4 are still given in terms of the solutions of Painleve III type equation that appeared in the study of $\C^1/\Z_3$ with different boundary conditions. For $\C^2/\Z_5$ case, governing equations are of generalized Toda type. The potential is monotonically decreasing function of RG flow.
\end{abstract}

\vskip 5cm
pacs:  {11.25.Mj}

\renewcommand{\theequation}{\thesection.\arabic{equation}}

\newpage


\section{Introduction}

\setcounter{equation}{0}
 
The study of localized tachyon condensation  \cite{aps,vafa,hkmm,dv,many} has been 
considered with  many interesting developments. 
The basic picture is that 
tachyon condensation induces cascade of decays of the orbifolds to  less singular ones until the spacetime 
supersymmetry is restored. Therefore the localized tachyon condensation has a geometric description as the resolution 
of the  spacetime singularities.  

Following the line of Vafa's reformulation of the problem in terms of Mirror Landau-Ginzburg theory,  we worked out the detailed analysis on the fate of spectrum and the background geometry under the tachyon condensation as well as the question of what is the analogue of c-theorem with the GSO-projection  
in a series of papers\cite{sin,sinlee,gso}. In all these works,  super-conformal invariance was used very heavily so  that we were working at the string theory on-shell level. 

On the other hand, recently, Dabholkar and Vafa\cite{dv} proposed that the tachyon  potential is  given by the maximal charge.
Strictly speaking, the $U(1)$ charge in the question  is defined only on the conformal point not on the off-shell.
On the other hand, the decay process considered as a renormalization group flow is  an off-shell process and one needs to extend the concept of charge in order to define the tachyon potential. The way to extend the charge is to consider it as a semi-index which is contributed by BPS objects only \cite{cvi}. These are not just topological since it has anti-holomorphic dependence as well as holomorphic dependence on the deformation parameters. The way to calculate this quantity is to show that it satisfies $tt^*$ equations\cite{cv} and solve it if possible.  In \cite{dv}, the process where $\C/\Z_3$ decays to  $\C^1$ was discussed. 

In this paper we extend the result of \cite{dv} to $\C^2/\Z_n$  for $n=3,4,5$. Interestingly, the tachyon potential for $n=3$ and 4 is still given in terms of the solutions of Painleve III type 
that appeared in the study of $\C^1/\Z_3$ with different boundary conditions. For $\C^2/\Z_5$ the governing equation is that of generalized Toda type. The potential is monotonically decreasing function of RG flow, therefore expected so as function of real time as well. For LG model associated to $\C^2/\Z_5$ without orbifold projection, $tt^*$ equation contains the Bullough-Dodd equation, whose solution is known\cite{kit}.

In this letter, we do not attempt to give all necessary background. For the set up of tachyon condensation in terms of mirror symmetry, see \cite{vafa,sin}. For the application of $tt^*$ to the tachyon condensation, see \cite{dv}.

\section{$\C^2/\Z_3$}
\setcounter{equation}{0}
In this section we study the tachyon condensation in $\C^2/\Z_n$ with Landau-Ginzburg (LG) description by calculating and solving the  corresponding $tt^*$ equations. 
First we consider the simplest of all $\C^2/\Z_n$, namely $n=3$.
The mirror of this is a LG whose  superpotential is given by
\be
W={x^3 \over 3} +{y^3 \over 3} - t x y,
\ee
with an imposition of an orbifold constraint. 


Before we consider the orbifolded LG theory, we work out the $tt^*$ equations for LG 
theory without orbifolding for later interests.  The fundamental variables are 
not $x,y$ but $\log x, \log y$ \cite{dv,hori} so that the chiral ring consists of 
\be
\{ x y, x^2y,xy^2,x^2y^2 \}.
\ee
from which charges of the elements can be read off to  give $NS$-charges  
\be
\{{2 \over 3},1,1,{4\over 3} \}.\ee
The R-charges in the topological strings are related to those in $NS$ by the spectral flow
\be
q_R=q_{NS}-{n\over 2}.
\ee
For $\C^2/\Z_3$, we have $n=2$ so we get 
\be
\{-{1\over 3},0,0,{1\over 3} \} .\label{charge1}\ee

This superpotential has symmetries 
\ba
&&x\to \omega x,~~y\to \omega^2 y ; \no
{\rm and } &&x\to \omega^2 x,~~y\to \omega y,
\ea
where $\omega=e^{{2\over 3}\pi i}$, which constrains the metric 
$g_{ {\bar j}i}:=\la \phi_{\bar j}\phi_i \ra (\phi_is~ {\rm are~ chiral 
fields})$ 
to be of the form
\be
g=\pmatrix { a_{11} & 0 & 0 & \bar{b} \cr
            0 & a_{21}& 0 & 0 \cr
          0 & 0 & a_{12} & 0 \cr
          b & 0 & 0 & a_{22}
}.
\ee
The topological metric $\eta_{ij}$ is given by residue paring 
$$
\eta_{ij}=\la\phi_i\phi_j\ra={1\over {(2\pi i)}^n}\int_\Gamma
{ {\phi_i(X)\phi_j(X)dX^1\wedge\cdots\wedge dX^n} \over 
{\partial_1 W \partial_2W\cdots \partial_nW} 
}~~~~~~
$$
with  superpotential $W$, which in this case is  calculated to be
\be
\eta=\pmatrix{ 0 & 0 & 0 & 1 \cr
         0 & 0 & 1 & 0 \cr
          0 & 1 & 0 & 0 \cr
          1 & 0 & 0 & t^2
}.
\ee

From the reality constraint
\be
\eta^{-1} g (\eta^{-1} g)^* =I,
\ee
we have
\be
b={t^2 \over 2}a_{11}, ~~a_{12}={1\over a_{21}}, ~~a_{22}={ 1\over a_{11}} +{|t|^2 \over
4}a_{11}. \label{exch1}
\ee
The chiral ring coefficient are defined by $\phi_i \phi_j=C^k_{ij}\phi_k$ and if we denote the  matrix  for the multiplication by $xy$ by $C_t$, then  
\be
C_t=\pmatrix { 0 & 0 & 0 & 1 \cr
          0 & t^2 & 0 & 0 \cr
          0 & 0 & t^2 & 0 \cr
         0 & 0 & 0 & t^2
}.
\ee
Putting all these into $tt^*$ equation
\be
\partial_{\bar{t}} (g \partial_{t} g^{-1})
=[C_t,g(C_t)^\dagger g^{-1}],
\ee  
we get 
\ba
&& \partial_{\bar{t}} \partial_t \log a_{11} =-{1\over a_{11}^2} +
{ |t|^8 \over 16} a_{11}^2, \no
&&\partial_{\bar{t}} \partial_t \log a_{12} =0 .\label{ttbar1}
\ea
The exchange symmetry 
$x \rightleftarrows y$ due to the special form of perturbation, 
together with the reality condition eq.(\ref{exch1}), determines 
\be
a_{12}=a_{21}=1,
\ee
which is consistent with the second equation of eq.(\ref{ttbar1}).
By changes of variables
\be
y=a_{11}^{-1}, ~~ \zeta={1\over 3}t^{3},~~{\rm and}~~ y={1\over
2}|t|^{2}Y, \label{defY}
\ee
the first equation of eq.(\ref{ttbar1}) can be transformed into 
\be
 \partial_{\bar{\zeta}} \partial_\zeta \log Y =\frac{1}{4}\( Y^{2} - Y^{-2}\) .
\label{formY} \ee
In terms of $Y$ and $z=|\zeta|$ the last equation can be written as
\be
Y''= {(Y')^2 \over Y} -{Y'\over z} +Y^3 -{1\over Y},
\ee
which is already known as Painleve III equation.
We can further rewrite the eq.(\ref{formY}) as the
sinh-Gordon equation
\be
 \partial_{\bar{\zeta}} \partial_\zeta u = \sinh u , \label{sinh}
\ee
by introducing $u$ by $u=2 \log y$.
Now, since we are interested in the scaling behavior of the solutions,
we look at the dependence on $z=|\zeta|$ of the eq.(\ref{sinh}),
\be 
u''+u'/z = 4 \sinh u , \label{sinhz}\ee
where the prime denote the derivative in $z$.
The solution to this is well known\cite{its}. 
The real solutions are classified by their asymptotic behavior in $z\to 0$
\be
u(z) \simeq r \log z + O(z^{2-|r|}) ~for ~|r|<2.
\ee
In our case, the regularity of $a_{11}$ requires  
\be u(z) \sim -{4\over3} \log z. \label{asymp0}
\ee
 Therefore this fixes $r=-4/3$. For $z\to \infty$
\be
u(z) \sim \sqrt{3\over \pi} { e^{-2z} \over \sqrt{z} } ,~~z\to \infty .\label{asympinfty}
\ee
The precise form of regular solution can be written in terms of a convergent expansion 
\be
u(z;r)=-4\sum^\infty_{n=0} {[2\cos((2-r)\pi/4)]^{2n+1}] \over
{2n+1}} \int^\infty_{-\infty} \prod^{2n+1}_{i=1}{d\theta_i \over {4\pi}}
{e^{-2z\cosh\theta_i }\over {\cosh[(\theta_i-\theta_{i+1})/2] } } \label{exactu}
\ee
with $\theta_{2n+1}\equiv \theta_1$ and $r=-4/3$.

In order to define the charge matrix, we need to look at the scaling behavior of the system. 
The scaling 
 $z \to \lambda z $  induces 
$
\int d^2z d^2 \theta  \to \lambda \int d^2z d^2 \theta  $, 
which is equivalent to the field redefinition and coupling change such that  $ W \to \lambda W$.
For the given superpotential
\be
W={x^3 \over 3} +{y^3 \over 3} -t xy,
\ee
we can identify the necessary field redefinitions and coupling change as  
\be
x= \lambda^{1\over
3}\tilde{x},~ 
y= \lambda^{1\over 3}\tilde{y}~~ {\rm and }~~
t = \lambda^{1\over 3}t_0, \label{scalet}
\ee
 by which  $W$ can be written as
\be
W= \lambda({\tilde{x}^3 \over 3} + {\tilde{y}^3 \over 3} -t_0 \tilde{x}\tilde{y}).
\ee
In this way changing $t$ is equivalent to the changing the scale. The metric components 
in old and new basis are related by
\be
a_{ij}=\la x^{\bar i}y^{\bar j}|x^iy^j \ra =|\lambda|^{(i+j)\over n} \la  
\tilde{x}^{\bar i}\tilde{y}^{\bar j}|\tilde{x}^i\tilde{y}^j \ra :=|\lambda|^{(i+j)\over n} b_{ij} ,\label{anb}
\ee
with $n=3$.
The off-shell `charge' of the system was defined \cite{cv} as
\be
Q=g\partial_\tau g^{-1} -{{\hat c}\over 2} \label{Q}
\ee
where $\tau=\log \lambda$ and the metric components in use are in $\tilde{x},\tilde{y}$ basis, 
namely $b_{ij}$'s.
From the regularity condition of 
the metric in $x,y$ basis at $t=0,$ one can verify that the charge at the
starting point $t=\lambda=0$ is encoded correctly  to give the result listed 
in eq.(\ref{charge1}). 

{\it Orbifolded LG model: $\{W=x^3+y^3-txy\}//\Z_3 $}\\
In this  case, the chiral ring generated by $xy$: $\{ xy, x^2y^2 \}$ and  $NS$-charges are  
$\{{2 \over 3},{4\over 3} \}$.
 The metric and charges can be obtained from the previous subsection by simply discarding 
$a_{12}$ and $a_{21}$.
Following \cite{dv}, the tachyon potential is proposed to be
\be
V(t,{\bar t})=2 {\rm max}|Q(t,{\bar t})|= -\frac{1}{2}z\frac{du}{dz},
\ee
where we  identified $\zeta=\lambda$ by eqs.(\ref{defY}),(\ref{scalet}) after setting 
$t_0=3^{1/3}$.
The factor 2 in the above equation is from 
the asymptotic form of $u$ is given by eqs. (\ref{asympinfty}),(\ref{asymp0}). 
Its exact form can be written 
in terms of $u$ given in eq.(\ref{exactu}) with $r=-4/3$. 
The potential vanishes exponentially as $t\to\infty$.
As a consequence of the tachyon condensation, the fate of the $\C^2/\Z_3$ is just $\C^2$ 
as expected.


\section{$\C^2/\Z_4$}

 
Let us now consider $n=4$. 
Working in the basis with definite ordering the chiral ring consists of
\be
\{xy,x^2y,x^3y,xy^2,x^2y^2,x^3y^2,xy^3,x^2y^3,x^3y^3  \},
\ee 
The topological metric and metric from sec.3.1 are given by 
\be
\eta_{x^{i_1}y^{j_1},x^{i_2}y^{j_2}}=1,~{\rm if}~ i_1+i_2=j_1+j_2=4,
~~~\eta_{x^3y^3,x^3y^3}=t
\ee
and 
\be
g_{i\bar{j}}= \pmatrix{ a_{11} & 0 & 0 & 0 & 0 & 0 & 0 & 0 & \bar{b}
\cr 0 & a_{21} & 0 & 0 & 0 & 0 & 0 & 0 & 0 \cr 0 & 0 & a_{31} & 0 & 0 &
0 & 0 & 0 & 0  \cr 0 & 0 & 0 & a_{12} & 0 & 0 & 0 & 0 & 0  \cr 0 & 0
& 0 & 0 & a_{22} & 0 & 0 & 0 & 0  \cr 0 & 0 & 0 & 0 & 0 & a_{32} & 0 & 0
& 0 \cr 0 & 0 & 0 & 0 & 0 & 0 & a_{13} & 0 & 0 \cr 0 & 0 & 0 & 0 & 0
& 0 & 0 & a_{23} & 0  \cr b & 0 & 0 & 0 & 0 & 0 & 0 & 0 & a_{33} }.
\ee
Then, the reality condition gives
\be
a_{22}=1,~a_{32}=1/a_{12},~a_{13}=1/a_{31},~a_{23}=1/a_{21},~a_{33}=1/a_{11} +|t|^4a_{11}/4,~b=t^2 a_{11}/4.
\ee
$C_t$ can also be calculated easily and given by
\be
C_t=\pmatrix{ 0 & 0 & 0 & 0 & 1 & 0 & 0 & 0 & 0 \cr 0 & 0 & 0 & 0
& 0 & 1 & 0 & 0 & 0 \cr 0 & 0 & 0 & 0 & 0 & 0 & t & 0 & 0 \cr 0 &
0 & 0 & 0 & 0 & 0 & 0 & 1 & 0 \cr 0 & 0 & 0 & 0 & 0 & 0 & 0 & 0 &
1 \cr 0 & t^2 & 0 & 0 & 0 & 0 & 0 & 0 & 0 \cr 0 & 0 & t & 0 & 0 &
0 & 0 & 0 & 0 \cr 0 & 0 & 0 & t^2 & 0 & 0 & 0 & 0 & 0 \cr 0 & 0 &
0 & 0 & t^2 & 0 & 0 & 0 & 0 }.
\ee
Then the $tt^*$ equation becomes
\ba
&&\label{1} ~~-\partial_{\bar{t}}\partial_t\log a_{11}=1/a_{11} 
-|t|^4a_{11}/4, \\
&&\label{2} ~~-\partial_{\bar{t}}\partial_t\log
a_{21}=1/(a_{21}a_{12})-|t|^4a_{21}a_{12}, \\
&&\label{3} ~~-\partial_{\bar{t}}\partial_t\log a_{31}=|t|^2/a^2_{31}
-|t|^2a^2_{31}, \\
 &&\label{4} ~~-\partial_{\bar{t}}\partial_t \log
a_{12}=1/(a_{21}a_{12})-|t|^4a_{21}a_{12}. \ea

Most of the equations are of the form
\be
\partial_{\bar{t}}\partial_t
\log y =y^2 -{|t|^{2k} \over m^2}y^{-2}. \label{standard}
\ee
By introducing change of variables
\ba
\zeta &=&(1/(1+k/2))(16/m^2)^{1/(2k+4)}t^{1+k/2},\cr
y&=&\sqrt{1/m}|t|^{k/2}e^{u/2}, \ea
above equation lead us  to the sinh-Gordon equation eq.(\ref{sinh}).
The value of the $r$ in the solution of sinh-Gordon equation 
can be obtained from the regularity of the metric component near $t=0$. 
Since $y\sim z^{k/k+2}e^{u/2}$, we have
\be
u\sim -\frac{2k}{k+2}\log z + \cdots, \label{u0}
\ee
which determines the value  $r=-\frac{2k}{k+2}$.

Now from eq.(\ref{Q}) the components of charge matrix is given by \be 
q_{ij}=b_{ij}\d_\tau b^{-1}_{ij}-1.
\ee
Using $a_{ij}=|\lambda|^{(i+j)/2}b_{ij}$, ~~ $a^{-1}_{ij}\sim y^l$ for some $l$ and  $y\sim z^{k/(k+2)}e^{u/2}$ with identification $z=|\lambda|$,
 \be
 q_{ij}(z)=\frac{l}{4}z\frac{du(z)}{dz}+\frac{lk}{2(k+2)} +\frac{i+j}{n}-1.
 \ee
 Notice that $(ij)$ is not the matrix index but the vector index.  
 Using eq.(\ref{u0}), the  value of the charge at $t=0$ is
 \be
 q_{ij}(0)= \frac{i+j}{n}-1,
 \ee
which confirms that we are in the right track.
Now we apply this result to our system.

For $a_{11}$, by change of variables
$a_{11}^{-1}=2y^2 $ the eq.(\ref{1})  reduces to the standard form eq.(\ref{standard}) with $k=2,~ m=4$. Therefore $r=-1$. 
This equation further can be reduced to sinh-Gordon equation
$\partial_{\bar{\zeta}}\partial_\zeta u=\sinh u$ by setting $\zeta=t^2/2$ and $y=|t|e^{u/2}/{2}$. Since $l=2$ in this case,  charge is given by 
$q_{11}(z)=\frac{1}{2}z\frac{du(z)}{dz}$.

For $a_{12}$ and $a_{21}$, first we show they are equal. From eq.(\ref{2}) and eq.(\ref{4}) $a_{12}=|F(t)|^2a_{21}$ for some holomorphic function $F(t)$. Since  $a_{12}=a_{21}$ at $t=0$ as well as at $t=\infty$, the only non-singular holomorphic function with such boundary conditions is a constant function $F(t)=1$, i.e, $a_{12}=a_{21}$. This supports the exchange symmetry $a_{ij}=a_{ji}$. 
With this, 
the eq.(\ref{2},\ref{4}) are the case of $m=1, k=2$; by setting
 $a_{12}^{-1}=y=|t|e^{u/2}$ and $\zeta=t^2/2$, we get sinh-Gordon
equation.  $l=1$ lead us to 
$q_{12}(z)=q_{21}(z)=\frac{1}{4}z\frac{du(z)}{dz}$.

For $a_{31}$, by $z=t^2$ and $y=a_3^{-1}$
we get sinh-Gordon and the solution is $y=e^{u/2}$.  It is easy to see $q_{31}(z)=\frac{1}{4}z\frac{du(z)}{dz}$. Notice that $q_{31}(0)=q_{13}(0)=0$. The monotonicity of the charge in $t$ suggests that $q_{31}(z)=0$. In fact  the exchange symmetry  $a_{31}=a_{31}$ and the reality condition $a_{31}=1/a_{31}$ sets $a_{31}=1$. 

{\it Mirror of $\C^2/\Z_4$: } 

So far, we have been considering the LG model without orbifolding action. To consider the mirror of $\C^2/\Z_4$ with generator $xy$, we need to consider $a_{ii}$ $i=1,2,3$. Since  $a_{22}=1$ and $a_{33}$ is given by $a_{11}$, we only need to consider the equation for $a_{11}$, which is given by eq.(\ref{1}). 


\section{$\C^2/\Z_5$}\setcounter{equation}{0}

Here again, we first analyze the general LG model and at the end we comments on the orbifolded case. 
The superpotential is given by
\be
W={x^5 \over  5} + {y^5 \over 5} -txy.
\ee
The chiral ring is give by $x^4-ty=0$ and $y^4-tx=0$. We order the
basis by dictionary order in charge pair $(i,j)$:
\ba
&&{\cal R} =\{ xy, x^2y, x^3y, x^4y, xy^2, x^2y^2, x^3y^2, x^4 y^2, 
xy^3, \no 
&&~~~~~~~~ x^2 y^3, x^3 y^3, x^4 y^3, xy^4, x^2 y^4, x^3 y^4, x^4 y^4 
\}.
\ea
The $\eta_{ij}$ can be readily written and we avoid to writing it down. 
The metric $g_{i\bar{j}}$ has 16 diagonal components which we denote 
by $a_{ij}=\bra{x^{\bar i}y^{\bar j}}|{x^iy^j}\ket, ~i,j=1,2,3,4$,
and two non-vanishing off diagonal elements $b, \bar b$ as before.
The coupling matrix $C_t$  in this basis is given by 
\be
C_t=\pmatrix{ 0 & 0 & 0 & 0 & 0 & 1 & 0 & 0 & 0 & 0 & 0 & 0 & 0 &
0 & 0 & 0 \cr 0 & 0 & 0 & 0 & 0 & 0 & 1 & 0 & 0 & 0 & 0 & 0 & 0 &
0 & 0 & 0 \cr 0 & 0 & 0 & 0 & 0 & 0 & 0 & 1 & 0 & 0 & 0 & 0 & 0 &
0 & 0 & 0 \cr 0 & 0 & 0 & 0 & 0 & 0 & 0 & 0 & t & 0 & 0 & 0 & 0 &
0 & 0 & 0 \cr 0 & 0 & 0 & 0 & 0 & 0 & 0 & 0 & 0 & 1 & 0 & 0 & 0 &
0 & 0 & 0 \cr 0 & 0 & 0 & 0 & 0 & 0 & 0 & 0 & 0 & 0 & 1 & 0 & 0 &
0 & 0 & 0 \cr 0 & 0 & 0 & 0 & 0 & 0 & 0 & 0 & 0 & 0 & 0 & 1 & 0 &
0 & 0 & 0 \cr 0 & 0 & 0 & 0 & 0 & 0 & 0 & 0 & 0 & 0 & 0 & 0 & t &
0 & 0 & 0 \cr 0 & 0 & 0 & 0 & 0 & 0 & 0 & 0 & 0 & 0 & 0 & 0 & 0 &
1 & 0 & 0 \cr 0 & 0 & 0 & 0 & 0 & 0 & 0 & 0 & 0 & 0 & 0 & 0 & 0 &
0 & 1 & 0 \cr 0 & 0 & 0 & 0 & 0 & 0 & 0 & 0 & 0 & 0 & 0 & 0 & 0 &
0 & 0 & 1 \cr 0 & t^2 & 0 & 0 & 0 & 0 & 0 & 0 & 0 & 0 & 0 & 0 & 0
& 0 & 0 & 0 \cr 0 & 0 & t & 0 & 0 & 0 & 0 & 0 & 0 & 0 & 0 & 0 & 0
& 0 & 0 & 0 \cr 0 & 0 & 0 & t & 0 & 0 & 0 & 0 & 0 & 0 & 0 & 0 & 0
& 0 & 0 & 0 \cr 0 & 0 & 0 & 0 & t^2 & 0 & 0 & 0 & 0 & 0 & 0 & 0 &
0 & 0 & 0 & 0 \cr 0 & 0 & 0 & 0 & 0 & t^2 & 0 & 0 & 0 & 0 & 0 & 0
& 0 & 0 & 0 & 0
 }.
\ee
By use of the reality condition,  we have 8 independent variables
out of 16+2 real variables and the rests can be written in terms of them:
\ba
a_{13}&=&1/a_{42},~a_{23}=1/a_{32},~a_{33}=1/a_{22},~a_{43}=1/a_{12},
~a_{14}=1/a_{41}, 
\no
a_{24}&=&1/a_{31},~a_{34}=1/a_{21},~a_{44}=1/a_{11}+|t|^4a_{11}/4,\no
b&=&\half t^2a_{11}.
\ea

Then the $tt^*$ equation becomes
\ba
&&-\partial_{\bar{t}}\partial_t \log a_{11} = a_{11}^{-1}a_{22} -{1\over
4}|t|^4 a_{11} a_{22}, \no
&&-\partial_{\bar{t}}\partial_t \log a_{22}= a_{22}^{-2} 
-a_{11}^{-1}a_{22}
-{1\over 4} |t|^4 a_{11} a_{22} \label{a1122},\\
&&-\partial_{\bar{t}}\partial_t \log a_{21}=
-|t|^4 a_{21} a_{12} + a_{21}^{-1}a_{32}, \no 
&&-\partial_{\bar{t}}\partial_t \log
a_{12}=-|t|^4a_{21} a_{12} + a_{12}^{-1}a_{32}^{-1}, \no
&&-\partial_{\bar{t}}\partial_t \log
a_{32}= a_{12}^{-1}a_{32}^{-1} -a_{21}^{-1}a_{32}, \no
&&-\partial_{\bar{t}}\partial_t
\log a_{31} = -|t|^2 a_{31}a_{41} + a_{31}^{-1}a_{42}, \no
&&-\partial_{\bar{t}}\partial_t \log a_{41}= -|t|^2a_{31} a_{41} +|t|^2
a_{41}^{-1}a_{42}^{-1}, \no 
&&-\partial_{\bar{t}}\partial_t \log a_{42}= |t|^2
a_{41}^{-1}a_{42}^{-1}-a_{31}^{-1}a_{42}.
\ea
The reality condition together with exchange 
symmetry gives us only 4 independent equations 
\ba
&&-\partial_{\bar{t}}\partial_t\log a_{11}=a_{22}/a_{11}-a_{11} a_{22} 
|t|^4/4, \label{11} \\
&&-\partial_{\bar{t}}\partial_t \log a_{22} =a_{22}^{-2}-a_{22}/a_{11} 
-|t|^4
a_{11}a_{22}/4, \label{22}\\
&&-\partial_{\bar{t}}\partial_t \log a_{21} =-|t|^4 a_{21}^2 + 
a_{21}^{-1}, \label{21}\\
&&-\partial_{\bar{t}}\partial_t \log a_{31} =-|t|^2 a_{31} + a_{31}^{-2}, 
\label{31}
\ea
as well as the predetermined values of some of them.  
 \be
  a_{32}=a_{23}=1, ~~  a_{41}=a_{14}=1.\ee
Notice that all monomials 
$x^2y, xy^2, x^3, y^3$  have NS charge 1 and these are the marginal operators. Above results show that the marginal operators do not evolve under the  condensation of tachyon represented by $xy$. 

To eliminate $t$ from above equations, let 
\be
a_{ij}:=|t|^{c_{ij}} b_{ij}, ~~~\zeta:=a t^b.\ee
Then by using  $\partial_{\bar{t}}\partial_t=
(ab)^2|t|^{2b-2}\partial_{\bar{\zeta}}\partial_\zeta$,
and by requiring that the eqs. (\ref{11}), (\ref{22}) are homogenous in $t$, 
\ba
2b-2&=&c_{22}-c_{11}=4+c_{22}+c_{11}, \no  &=&-2c_{22}=-c_{11}+c_{22}, \ea
which give $c_{22}=-2/3$, $c_{11}=-2$ and $b=5/3$ with $ab=1$. 
Based on this, we introduce $q_{11}$, $q_{22}$ by 
\be
a_{11}=2(3/5)^{2}|\zeta|^{-6/5}e^{-q_{11}}, 
~~~a_{22}=2^{1/3}(3/5)^{2/3}|\zeta|^{-2/5}e^{-q_{22}}, \label{a0011}
\ee
and  re-scale  by $\zeta\to  {2}^{1/3}\zeta$ to get
\ba
&&\partial_{\bar{\zeta}}\partial_\zeta 
q_{11}=e^{q_{11}-q_{22}}-e^{-(q_{11}+q_{22})}, \no
&&\partial_{\bar{\zeta}}\partial_\zeta 
q_{22}=e^{2q_{22}}-e^{q_{11}-q_{22}}-e^{-(q_{11}+q_{22})}. \label{q1122}
\ea

For eqs.(\ref{21}) and (\ref{31}), 
\be
4+2c_{21}=-c_{21}=2b-2=2+c_{31}=-2c_{31}.\ee Then we have
$c_{31}=-2/3$, $c_{21}=-4/3$ and $b=5/3$.  
We introduce $\tau$ and $Y(\tau), Z(\tau)$ by 
\be
\tau=|\zeta|^2,~~~ a_{21}=\({5}/{3}\)^{-4/5} \tau^{-2/5} e^{-Y} ~~{\rm 
and}~  
a_{31}=\({5}/{3}\)^{-2/5} \tau^{-1/5} e^Z .\label{a1020}\ee 
Then,  both eq.(\ref{21}) eq.(\ref{31}) are reduced to 
\be
\partial_\tau(\tau \partial_\tau Y)=e^Y-e^{-2Y},
\ee
\be
\partial_\tau(\tau \partial_\tau Z)=e^Z-e^{-2Z},
\ee
which are  known as Bullough-Dodd equation which is a degenerate 
Painleve III, which also appears in the case  $\C^1/\Z_4 \to \C^1$ 
transition with $W=x^4-tx$.

The  Bullough-Dodd equation
\be
\partial_\tau(\tau\partial_\tau u) =e^u -e^{-2u}
\ee
has been studied extensively and 
the properties of the asymptotically regular solutions 
were given  in  \cite{kit}. The solutions are parametrized
by four complex numbers $g_1$, $g_2$, $g_3$, and  $s$ satisfying
\be
g_1+g_2(1-s)+g_3=1,~~~g_2^2-g_1g_3=g_2.
\ee
The asymptotic forms are given by 
\ba
&&\tau\to \infty;~~e^u \sim 1+\sqrt{ {3\over \pi} } {s\over 2} 
(3\tau)^{-{1 \over 4}} e^{-2\sqrt{3\tau} }, ~~~g_1=g_2=0,~g_3=1, \no
&&\tau \to 0;~~e^u =-{ \mu^2 \over { 2\tau \sin^2 \{ {i\over 2} [\mu 
\ln\tau +\ln ( r_1{C_2\over C_0} ) ] \} } } \sim 2 
\mu^2 r_1 {C_2\over C_0} \tau^{\mu-1}, \no 
&&~~~~~~~~s=1+ \cos[{2\pi\mu \over 3}],~~~r_1=g_3-g_1 
+(1+\omega)(g_1-g_2),~~~\omega^\mp=e^{\mp{2\pi i \mu \over 3} },\no
&&~~~~~~~~{C_2 \over C_0}=3^{-2\mu} {\Gamma(1-{\mu\over 
3})\Gamma(1-{2\mu\over 3}) \over
{\Gamma(1+{\mu\over 3})\Gamma(1+{2\mu\over 3})} }.
\ea
Regularity of the metric as $\tau \to 0$ can fix $s$.

First let us consider $a_{21}\sim \tau^{-2/5}e^{-y}$.
From the regularity of $a_{21}$, we have 
\be
e^u=e^y\sim \tau^{-2/5},
\ee
which gives
\ba \mu=3/5, 
&&s=1+2\cos(2\pi/5) \simeq 1.618, ~~~r_1=1,\no
&&{C_2\over C_0}=3^{-6/5}(25/2){\Gamma(4/5)\Gamma(3/5) \over { 
\Gamma(1/5)\Gamma(2/5)} }.
\ea
Similarly, for $a_{31}\sim \tau^{-1/5}e^Z$, we have
\be
e^u=e^{-Z}\sim \tau^{-1/5},
\ee 
from which we get 
\ba
 \mu =4/5, &&s=1+\cos(8\pi/15)\simeq 0.791,~~~r_1=1, \no
&&{C_2\over C_0} =3^{-8/5}((15)^2/32)
{ \Gamma(11/15)\Gamma(7/15) \over {\Gamma(4/5)\Gamma(8/15) } }.
\ea
These completely fixes behaviors of solutions at both ends.

{\it Mirror of $\C^2/\Z_5$ :}

In this case, we only need to consider $a_{11},a_{22}$ since $a_{33}$ and $a_{44}$ are determined in terms of $a_{22}$ and $a_{11}$ respectively by the reality condition.
The $tt*$ equations for $a_{11},a_{22}$ are given by eqs.(\ref{a1122}) or eqs.(\ref{q1122}). 
They can be identified as the ${\tilde B}_2=D^T(SO(5))$ Toda system. We will investigate the relation between the $tt*$ equations in the orbifold geometry and various Toda systems elsewhere.
The charge matrix $Q=g\d_\tau g^{-1}-1$ with $\tau=\log\lambda$  can be calculated to be given by
\be  Q=\left(%
\begin{array}{cccc}
   -\frac{3}{5}+a_{11}\d_\tau a_{11}^{-1}  & 0 & 0 & 0 \\
  0&  -\frac{1}{5}+a_{22}\d_\tau a_{22}^{-1}  & 0 & 0 \\
  0 & 0 &  \frac{1}{5}-a_{22}\d_\tau a_{22}^{-1} & 0 \\
  t^2 a_{11}\d_\tau a_{11}^{-1}  & 0 & 0 &  \frac{3}{5}-a_{11}\d_\tau a_{11}^{-1}  \\
\end{array}%
\right). 
\ee
Notice that in terms of $q_{ij}$ and $\lambda(=\zeta)$, 
and if we look at the $|\lambda|$ dependence only, the tachyon potential can be identified as 
\be
V=2Q_{max}= -  \zeta\d_{\zeta}q_{11}(\zeta).
\ee
We expect that this is monotonically decreasing from the value $5/3$ at $t=0$ to $0$ at $t\to \infty$. So far, the mathematical literature on the solution to the equation is not available and the qualitative behavior we suggested above is from physical intuition that in the final stage of tachyon condensation there is no nontrivial chiral primaries with charge other than 0.  

\section{Discussion}

In this paper, we calculated $ tt^*$ equations for $\C^2/\Z_n$ with n=3,4,5. In $n=3,4$ cases, they reduce to a Painleve III equation with different boundary conditions. 
In $n=5$ case, they reduce to a simple Toda system whose explicit solutions are not known yet. 
Non-orbifolded LG model associated to $n=5$ case involves a Bullough-Dodd equation.

As a limitation of this paper, we mention that we considered the string theories without GSO projection only.  Since GSO projection does not provide a supersymmetry immediately in the orbifold background, there is not much point on restricting ourselves to GSO projected theory. 
According to the rule given in \cite{gso}, $xy$ term considered in this paper is projected out for type II, and we need to consider the deformation by higher operator, 
which result in highly non-trivial equations due to the algebraic complexity of reality condition.
Another very interesting case is the one where the daughter theory is also an orbifold. This also results in a highly non-trivial equations even for $\C^1/\Z_n$ background. We wish to report on these issues in later publications. 

\vskip .5cm 
\noindent {\bf \large Acknowledgement} \\
We want to thank J.Raeymaekers  H.Yee for discussions and  A. Sen for useful suggestions. This work is supported by the Korea Research Foundation Grant (KRF-2004-015-C00098).

\end{document}